\documentclass[prl,twocolumn]{revtex4}
\usepackage{amsmath}
\usepackage{epsfig}
\usepackage{float}
\usepackage{dcolumn} 

\begin{document}
\title{\bf Non-catastrophic resonant states in one dimensional scattering from a rising exponential potential}  
\author{Zafar Ahmed$^{1}$}
\affiliation{$^1$Nuclear Physics Division, Bhabha Atomic Research Centre, Mumbai, 400085, India} 
\author{Lakshmi Prakash$^{2}$}
\affiliation{$^2$University of Texas at Austin, Austin, TX, 78705, USA}
\author{Shashin Pavaskar$^{3}$}
 \affiliation{$^3$National Institute of Technology, Surathkal, Mangalore, 575025, India}
\email{1: zahmed@barc.gov.in, 2: lprakash@utexas.edu, 3: spshashin3@gmail.com} 
\date{\today}
\begin{abstract}
\noindent
Investigation of scattering from rising potentials has just begun, these unorthodox potentials have earlier gone unexplored. Here, we obtain reflection amplitude ($r(E)$) for scattering from a two-piece rising exponential potential: $V(x\le 0)=V_1[1-e^{-2x/a}], V(x > 0)=V_2[e^{2x/b}-1]$, where $V_{1,2}>0$. This potential is repulsive and rising for $x>0$; it is attractive and diverging (to $-\infty$) for $x<0$. The complex energy poles (${\cal E}_n= E_n-i\Gamma_n/2, \Gamma_n>0$)  of $r(E)$ manifest as resonances.  Wigner's reflection time-delay displays peaks  at energies $E(\approx E_n$) but the eigenstates do not show spatial catastrophe for $E={\cal E}_n$. 
\end{abstract}

\maketitle
In one dimension, quantum mechanical scattering is usually studied with three kinds of potentials ($V(x)$): (1) when $V(\pm \infty) \rightarrow {\cal C}$, (2) when $V(-\infty) \rightarrow {\cal C}$ and $V(\infty)\rightarrow -\infty$ or vice versa, and (3) when $V(\pm \infty)\rightarrow -\infty$, ${\cal C}$ being real and finite. Scattering from potentials that rise monotonically on one side has not been investigated until recently [1]. It is known that physical poles of transmission or reflection amplitudes $(r(E))$ give rise to possible bound and resonant states. Yet, one cannot
study scattering from all rising potentials even numerically due to the lack of mathematical amenability of Schr{\"o}dinger equation for a given potential even when $E$ is ignored  for asymptotically large distances. In this Letter, we derive exact analytic reflection amplitude for a rising exponential potential 
(see Fig. 1 and Eq. (2) below).
\begin{figure}[ht]
\centering
\includegraphics[trim = 0 0 0 16 cm, clip = true]{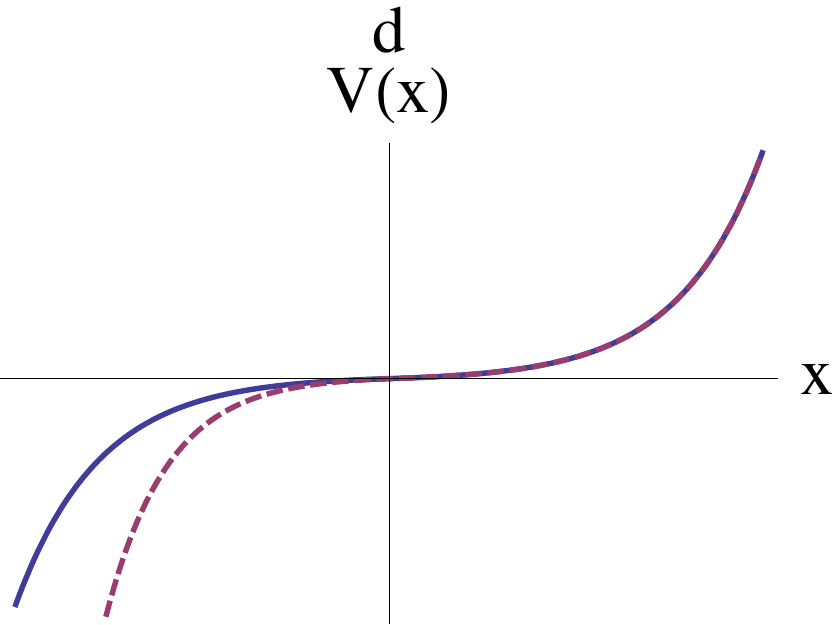}
\caption{Schematic depiction of rising exponential potential, $V(x)$ (Eq. (2)); solid line is anti-symmetric and the dashed line is asymmetric.}
\end{figure}

In the literature,  $V_{\lambda}=x^2/4-\lambda x^3$ [2] is available as a rising potential; this is not analytically solvable as a scattering or bound state problem. In textbooks, it is treated as a perturbed harmonic oscillator for finding discrete real bound states (though there are none). Nevertheless, by employing excellent approximate bound state methods, {\it non-real} discrete spectrum (${\cal E}_n=E_n-i\Gamma/2, \Gamma_n>0)$ has been found for this potential. These methods use outgoing wave (Gamow-Siegert) boundary condition [3] or complex scaling of coordinate [4]. Since this potential has three real turning points, it can ordinarily have discrete complex energy metastable states below/around the barrier of height $V_b$ ($E_n \sim V_b$). But the cause of  extraordinary presence of resonances well above the barrier, i.e.  $E_n>V_b$, has not been paid attention. We remark that this is due to rising part of the potential when $x \rightarrow -\infty$. Similarly, the complex discrete spectrum (resonances) of $V(x)=-x^3$ [5] found by interesting perturbation methods could not raise the issue of scattering from a rising potential. 

The issue of scattering from rising potential has commenced [1] due to a fine analytic derivation of the s-matrix, $S(E)$, for the odd parabolic potential, $V(x)=-x|x|$. It turns out that one can demand $\psi(x) \rightarrow 0$ for the side where the potential rises and on the other side one can seek a linear combination of incident and reflected wave as per the solution of Schr{\"o}dinger equation for the potential. One can then find the reflection amplitude, $r(E)=e^{i\theta(E)}$;
this is uni-modular as the rising potential would reflect totally. Nevertheless,  the Wigner's reflection time-delay [3]
\begin{equation}
\tau(E)=\hbar \frac{d\theta(E)}{dE},
\end{equation}
as a function of energy ($E$) can display maxima or peaks at $E=\epsilon_n$ with $\epsilon_n \approx E_n$. The discrete complex
energy eigenstates, called Gamow (1928) and Siegert (1939) states [3], decay time-wise but oscillate and grow asymptotically on one (both) side(s) of the potential if it has three (four) real turning points. This behavior of Gamow-Siegert states is well known as catastrophe in $\psi_n(x)$.
The scattering from a rising potential is essentially uni-directional, i.e, from left to right if the  potential rises for $x>0$ and vice-versa (see Fig. 1).

Curiously, the odd parabolic potential yields only one peak in $\tau(E)$ at $E\approx E_0$ and the eigenstate $\psi_{0}(x)$ at $E=E_0-i\Gamma_0/2$ shows no catastrophe.  We remark that  a single broad peak in time-delay (at $E=V_0$) is a common feature of many  single-top potential barriers as well, parabolic potential $V(x)=V_0-\frac{1}{2} m \omega^2 x^2$ [6],  Eckart barrier $V(x)=V_0 \mbox{sech}^2x$ [7], and Morse barrier [8] ($V_0>0$) being just a few simple examples.

Contrary to the above scenario, recently [9] two-piece rising potentials irrespective of number of real turning points have been shown  to possess metastable or resonant states 
displaying maxima or peaks in $\tau(E)$ at $E=\epsilon_n \approx E_n$, where ${\cal E}_n=E_n-i\Gamma_n/2,\Gamma_n>0$ are the poles of $r(E)$. $\psi_n(x)$ displays spatial catastrophe
for $E={\cal E}_n$ on the left ($x < 0$). These potentials consist of a rising part next to a step, a well, or a  barrier, or even a zero potential, i.e., $V(x)\rightarrow {\cal C}$  (real and finite) as $x \rightarrow -\infty)$.
One of the rising potentials considered in Ref. [9] is $V(x<0)=hx,
V(x\ge 0)=gx$; this diverges to $-\infty$  as $x\rightarrow -\infty$.
Remarkably, it displays the same qualitative features as others that converge at $x=-\infty$. Also, the absence of resonances in the smooth one-piece counterparts of these potentials (like Morse oscillator, linear and exponential) has been established. For rising part of the potential, linear, exponential and parabolic potentials have been used.  

Earlier, in an interesting detailed study [10] the rectangular and delta potentials in a semi-harmonic background have been found to possess complex discrete spectrum. However, the role of semi-harmonic (half-parabolic) potential has been unduly over emphasized. This however is one positive yet latent step towards studying  scattering from rising potentials.

In this Letter, we wish to present a two piece exponential potential which rises monotonically for $x\rightarrow \infty $ but unlike many potentials considered in Ref. [9] this potential diverges to $-\infty$ for $x\rightarrow -\infty$. This is a single real turning point unorthodox potential which entails parametric regimes displaying single broad to multiple sharp resonances. Eventually, we see that  this potential like the odd-parabolic 
rising potential  produces resonant states devoid of spatial
catastrophe.

We consider the one-dimensional potential:
\begin{equation} 
V(x)=\left\lbrace\begin{array}{lcr}
V_1[1-e^{-2x/a}], & &  x \le 0\\
V_2[e^{2x/b}-1], & & x >  0, \\
\end{array}
\right.
\end{equation}
where $a,b,V_1,V_2>0$.
For $x<0$ the time-independent  Schr{\"o}dinger equation can be written as
\begin{equation}
\frac{d^2\psi}{dx^2}+(p^2-s^2 e^{-2x/a})\psi(x),
\end{equation}
where $p=\sqrt{2m[E-V_1]}/\hbar$, $s=\sqrt{2mV_1}/\hbar$.
This equation can be transformed into cylindrical Bessel equation,
using $y=sae^{-x/a}$. Eq. (3) becomes
\begin{equation}
y^2\frac{d^2\psi(y)}{dy^2}+y\frac{d\psi(y)}{dy}+(y^2+p^2a^2)\psi(y)=0
\end{equation}
This Bessel equation [11] admits two sets of linearly independent solutions. For our purpose, we seek $H^{(1)}_{ipa}(y)$ and $H^{(2)}_{ipa}(y)$, where $p$ is allowed to  be real and purely imaginary. 
\begin{equation}
H^{(1,2)}_{\nu} (y)\sim \sqrt{2 /\pi y}~ e^{[\pm i(y-\nu \pi/2-\pi/4)]}, \quad ~ y \sim \infty. 
\end{equation}
Taking into account the time dependence, the total wave function
can be written as 
\begin{eqnarray}
\Psi_1(x,t)= H^{(1)}_{ipa}(y) ~e^{-iEt/\hbar} \nonumber \\ \sim  e^{ p \pi a/2} e^{i\pi/4} \sqrt{2/\pi y}~ e^{i[sae^{-x/a}-Et/\hbar]}.
\end{eqnarray}
As the time  increases ($t \rightarrow \infty$), to keep the phase $\phi=[sae^{-x/a}-Et/\hbar]$ constant, $x$ has to be large negative (away from $x=0$ on the left).
Therefore, $\Psi_1(x,t)$ represents a wave going from right to left after being reflected by the potential, $V(x)$ (2). Similarly,
\begin{eqnarray}
\Psi_2(x,t)=  H^{(2)}_{ipa}(y)~e^{-iEt/\hbar} \nonumber \\ \sim e^{-p \pi a/2} e^{-i\pi/4} \sqrt{2/\pi y} ~ e^{-i[sae^{-x/a}+Et/\hbar]}.
\end{eqnarray}
When  $t \rightarrow \infty$, to keep the phase $\chi=[sae^{-x/a}+Et/\hbar]$ constant,  $x$ being negative has to tend towards $x=0$. Thus, $\Psi_2(x,t)$ represents a wave traveling from left to right incident on the potential from left.
Eventually, for this potential (2)
\begin{equation}
\psi_i \sim e^{x/2a} e^{-isae^{-x/a}} \quad \mbox{and} \quad \psi_r \sim e^{x/2a} e^{isae^{-x/a}}
\end{equation}
act as asymptotic forms of incident and reflected waves similar to the plane waves $e^{\pm ikx}$. Probability  flux carried by the state $\psi(x)$ is defined as 
\begin{equation}
J=(2im\hbar)^{-1}\left(\psi^* \frac{d\psi}{dx}-\frac{d\psi^*}{dx} \psi\right).
\end{equation}
For $\psi_i$ and $\psi_r$, $J_i$ and $J_r$ are equal and opposite in sign as $s/(m\hbar)$ and $-s/(m\hbar)$ respectively.

Finally we seek the solution of (3) for (2) when $x<0$ as
\begin{eqnarray}
\psi(x)= A  e^{p \pi a/2} e^{-i\pi/4} H^{(2)}_{ipa}  (sae^{-x/a})\\ \nonumber + B   e^{-p \pi a/2} e^{i\pi/4}H^{(1)}_{ipa} (sae^{-x/a})
\end{eqnarray}
For $x>0$, by inserting the potential (2) in Schr{\"o}dinger equation and using the transformation: $z=ub e^{x/b}$, we get 
the modified cylindrical Bessel Equation given below.
\begin{equation}
z^2\frac{d^2\psi(z)}{dz^2}+z\frac{d\psi(z)}{dz}-(z^2+q^2a^2)\psi(z)=0, 
\end{equation}
where $q=\frac{\sqrt{2m(E+V_2)}}{\hbar}, u=\frac{\sqrt{2mV_2}}{\hbar}$.
This second order differential equation has two linearly independent solutions $I_{\nu}(z),K_{\nu}(z)$. Most importantly, we choose the latter noting that $K_{\nu}(z) \sim \sqrt{\frac{\pi}{2z}} e^{-z}$
when $z \sim \infty$.
So, we write
\begin{equation}
\psi(x) = C K_{iqb}(ube^{x/b}),\quad x>0.
\end{equation}
By matching $\psi(x)$ and its derivative at $x=0$ we get: 
\begin{eqnarray}
A e^{p \pi a/2} e^{-i\pi/4} H^{(2)}_{ipa}(sa) + B e^{-p \pi a/2} e^{i\pi/4}H^{(1)}_{ipa}(sa)\nonumber \\ = C K_{iqb} (ub)
\end{eqnarray}
and 
\begin{eqnarray}
A  e^{p \pi a/2} e^{-i\pi/4} {H^{(2)}_{ipa}}'  (sa) + B   e^{-p \pi a/2} e^{i\pi/4} {H^{(1)}_{ipa}}' (sa)\\ \nonumber = - \eta C K_{iqb}' (ub),\quad \eta=\sqrt{V_2/V_1}.
\end{eqnarray}
These two equations give 
\begin{small}
\begin{equation}
r(E)=B/A= ie^{p \pi a} \left( \frac{\eta K'_{iqb} (ub)H^{(2)}_{ipa}(sa)+
K_{iqb} (ub){H^{(2)}_{ipa}}'(sa)}{\eta K'_{iqb} (ub) H^{(1)}_{ipa}(sa)+K_{iqb}(ub) {H^{(1)}_{ipa}}'(sa)} \right).
\end{equation}
\end{small}
The modified Bessel function of second kind, i.e., $K_{i\nu}(z)$ is always real for real $\nu$ and $z$.
When $E<V_1$, $p$ becomes imaginary (say, $p=i\kappa$). Using the definition  of Hankel function we know that $H^{(1)}_{\nu}(z)=[H^{(2)}_{\nu}(z)]^*$ when $\nu$ and $z$ are real. Thus,  the unitarity of reflection  $|r(E<V_1)|=1$ follows. 
Further, using the properties $H^{(1)}_{-\nu}(z)= e^{i\nu \pi} H^{(1)}_{\nu}(z)$, $H^{(2)}_{-\nu}(z)= e^{-i\nu \pi} H^{(2)}_{\nu}(z)$
[11], we can absorb the factor $e^{p\pi a}$ and re-write $r(E)$ as
\begin{equation}
r(E)= i\left( \frac{\eta K'_{iqb} (ub)H^{(2)}_{-ipa}(sa)+K_{iqb}(ub) H^{(2)'}_{-ipa}(sa)}{\eta K'_{iqb} (ub) H^{(1)}_{ipa}(sa)+K_{iqb}(ub) {H^{(1)}_{ipa}}'(sa)} \right).
\end{equation}
Since the Hamiltonian is real Hermitian, unitarity must exist for  $E>V_1$ as well. We have derived an interesting property of the cylindrical Hankel function, namely
\begin{equation}
H^{(1)}_{i\nu}(z)=[H^{(2)}_{-i\nu}(z)]^*, \quad \nu ~\mbox{and} ~ z ~ \mbox{are real}
\end{equation}
which enables one to see that $R(E) = |r(E)|^2 = 1$. Equations (15,16) 
are actually identical, the latter performs (numerically) better
at higher energies.
The expressions of $r(E)$ in (15-16) are valid strictly for $V_1,V_2 \ne 0$; the cases of $V_1 = 0$ and $V_2 = 0$ need to be obtained separately. This is so because the limits $V_1\rightarrow 0$ $(V_2 \rightarrow 0$) and $x\rightarrow -\infty$ ($x\rightarrow \infty$)
do not commute. 

For the case $V_1=0$, we seek $\psi(x)=A e^{ikx} +B
e^{-ikx}$ for $x<0$, and $\psi(x)= K_{iqb}(ube^{x/b})$ for $x>0$, as the appropriate solution of Schr{\"o}dinger equation. Then, we get
\begin{equation}
r(E)=\frac{ik K_{iqb}(ub)-u K'_{iqb}(ub)}{ik K_{iqb}(ub)+u K'_{iqb}(ub)}.
\end{equation}
It can be readily checked that $|r(E)|^2=1$ holds for this rising potential as well.

For $V_2=0$, $V(x)$ is no more a rising potential; it is an orthodox potential allowing both reflection from it and transmission through it.
It is interesting to note that a particle incident on this potential
does not face a barrier in front in a  classical sense, yet it is both reflected and transmitted. It is useful to obtain $r(E)$ which will not be uni-modular. For $x \le 0$, we seek the same solution as in (10), but consider the plane wave, $\psi(x)=C e^{ikx}$, as solution for $x>0$. Then, we get 
\begin{equation}
r(E)=-e^{i\pi/2}~ e^{p \pi a} ~\frac{ikH^{(2)}_{ipa}(sa) + s H^{(2)'}_{ipa}(sa)}{ikH^{(1)}_{ipa}(sa)+s H^{(1)'}_{ipa}(sa)},
\end{equation}
the factor $e^{p \pi a}$ can again be absorbed in the numerator/denominator as done above in Eq. (16).
It can be checked that $|r(E<0)|^2 = 1$ (incident particle faces an infinitely thick barrier). Else, we get $|r(E)|^2<1$ (see the solid line in Fig. (11))
and this time we get non-zero transmittance $T(E)=1-|r(E)|^2$.
Earlier, reflection/transmission from the exponential potential
$V(x)=-V_0 e^{\pm x/c}$, $V_0>0$ has been studied [12]. According to this, $R(E)$ and $T(E)$ for $V(x)=V_1(1-e^{-2x/a})$ can be written as  $R(E>V_1) = e^{-\sqrt{(E+V_1)/\Delta}}$, $T(E>V_1)=1-e^{-\sqrt{(E+V_1)/\Delta}}$, where $\Delta={\hbar^2/(2ma^2)}$. 
Whereas, $R(E<V_1)=1$, $T(E<V_1)=0$ (see the dashed line in Fig. (11)). This scattering coefficient differs from Eq. (19) wherein the potential is two piece. 

\begin{figure}[H]
\centering
\includegraphics[trim = 0 0 0 16 cm, clip = true]{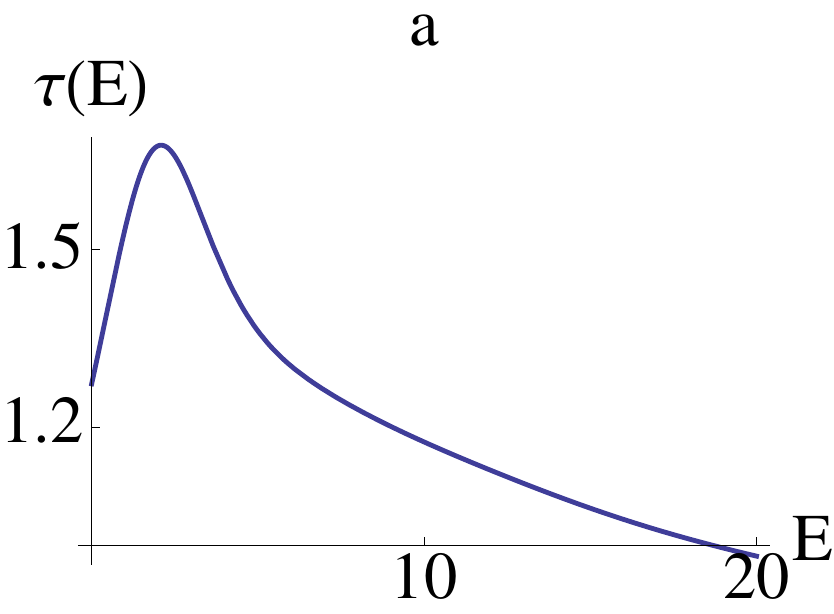}
\caption{A typical scenario of single peak in time-delay when the rising potential is anti-symmetric; see {\bf P1} in Table 1.}
\end{figure}

\begin{figure}[H]
\centering
\includegraphics[trim = 0 0 0 16 cm, clip = true]{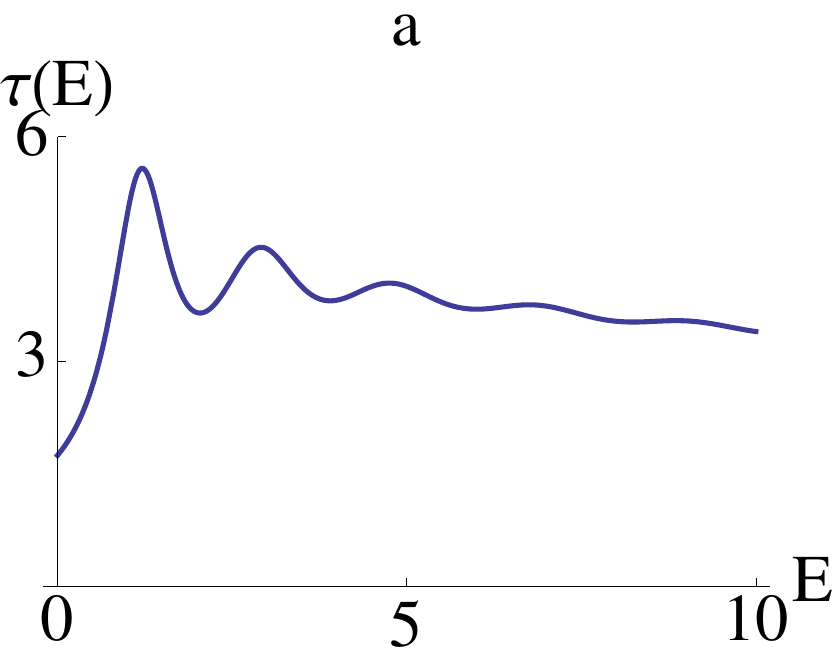}
\caption{Multiple maxima in time-delay for asymmetric case ({\bf P2}). See Table 1 for details.}
\end{figure}
\begin{figure}[H]
\centering
\includegraphics[trim = 0 0 0 16 cm, clip = true]{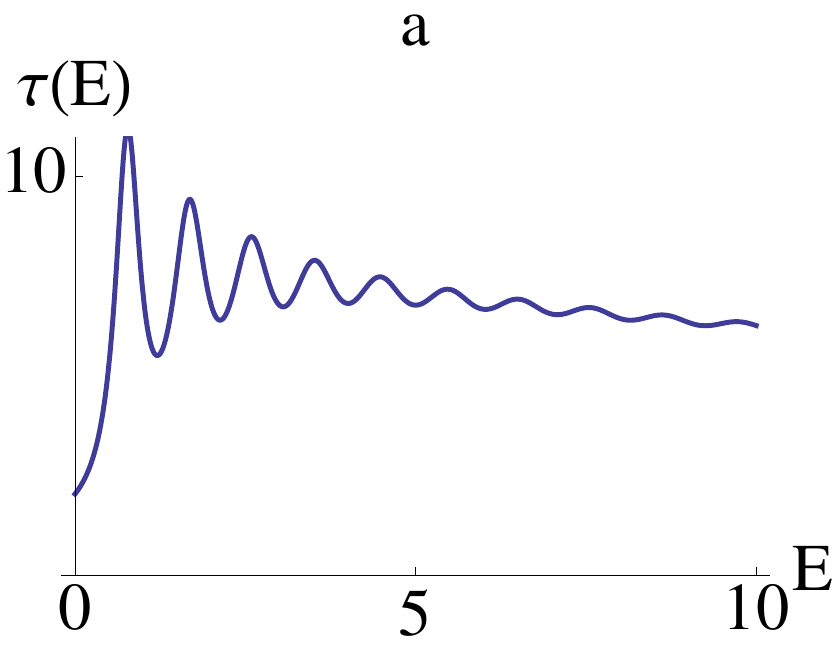}
\caption{Same as in Fig. 3 for {\bf P3}.}
\end{figure}

\begin{figure}[H]
\centering
\includegraphics[trim = 0 0 0 16 cm, clip = true]{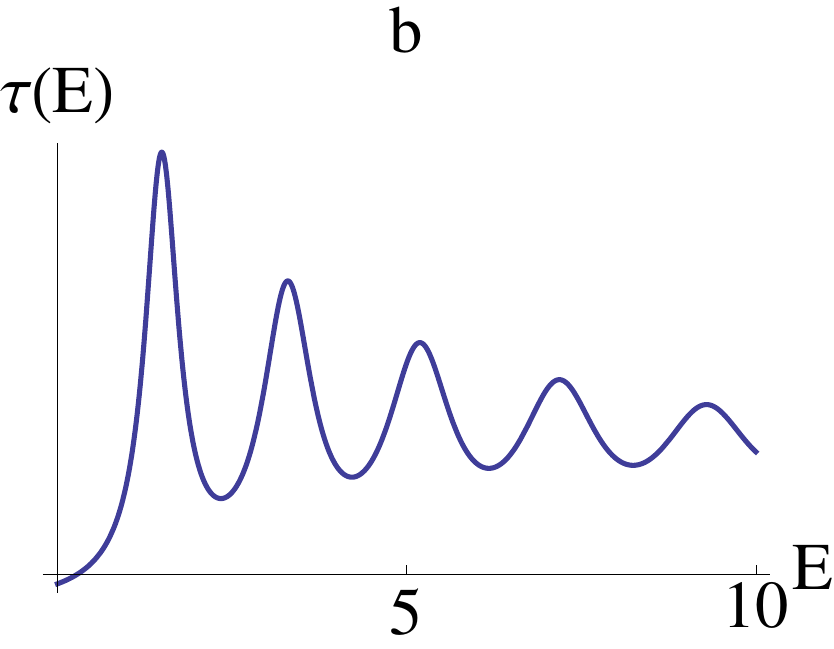}
\caption{Same as in Fig. 3 for {\bf P4}.}
\end{figure}

\begin{figure}[H]
\centering
\includegraphics[trim = 0 0 0 16 cm, clip = true]{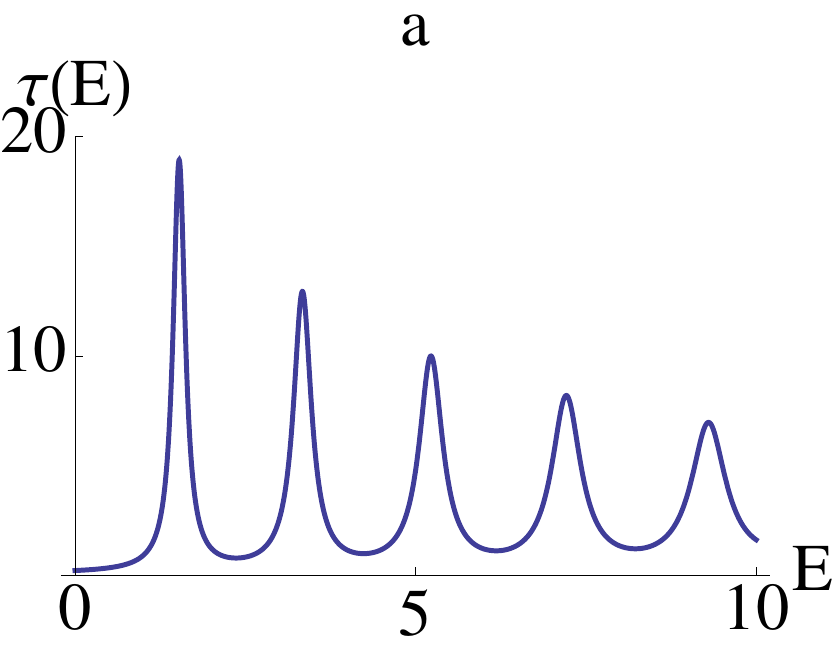}
\caption{Same as in Fig. 3 for {\bf P5}.}
\end{figure}
\begin{figure}[H]
\centering
\includegraphics{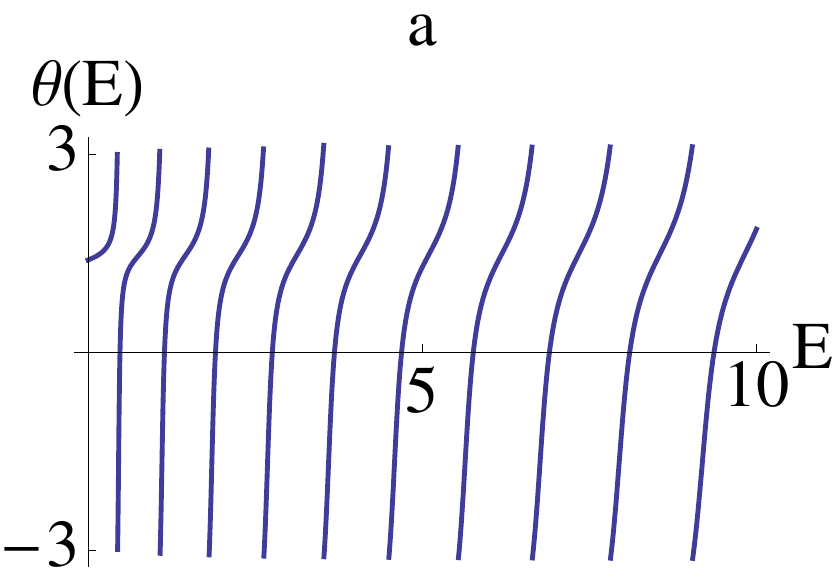}
\end{figure}
\begin{figure}[H]
\centering
\includegraphics{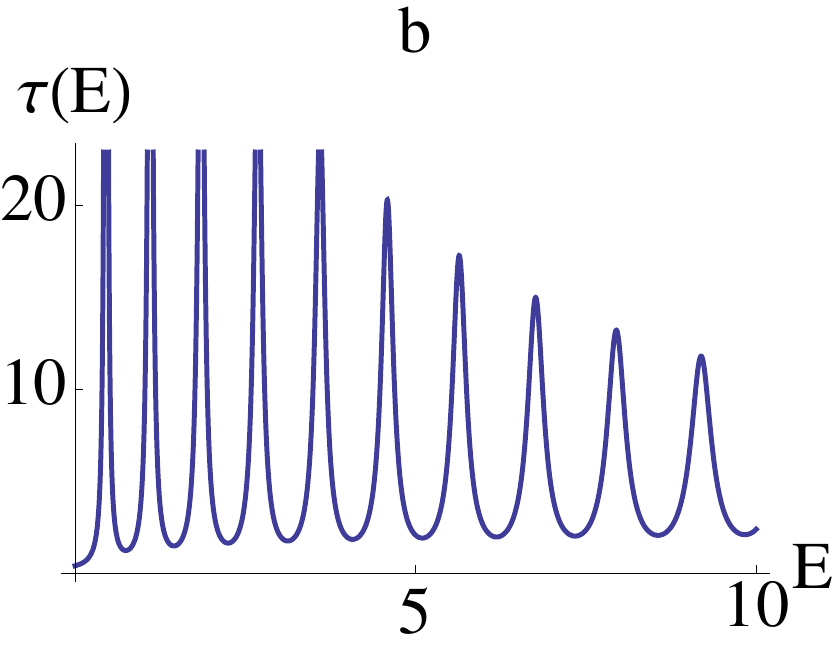}
\caption{Sharp resonances in {\bf P6}; a: phase shift ($\theta (E)$) and b: time-delay.}
\end{figure}
\begin{figure}[H]
\centering
\includegraphics{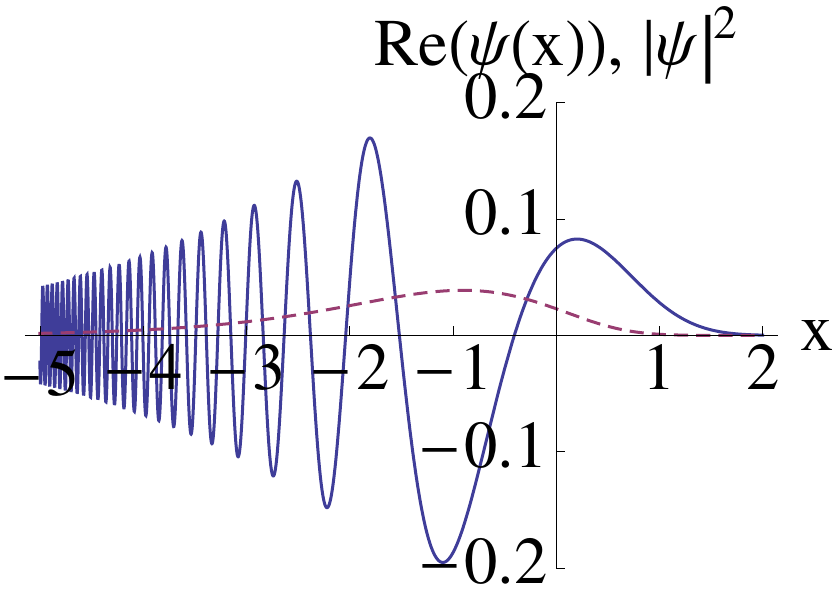}
\caption{Depiction of first resonant eigenstate in {\bf P1} at $E = 1.8305 - 2.4867 i$ : $\Re (\psi(x))$ (solid line) and $|\psi|^2$ (dashed line). Notice absence of catastrophe.}
\end{figure}
\begin{figure}[H]
\centering
\includegraphics[height = 5cm, width = 8cm]{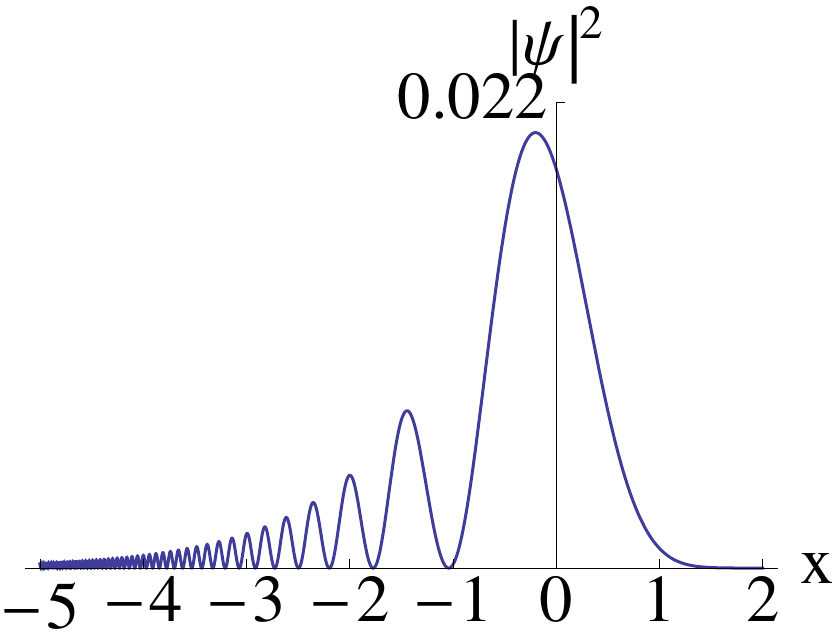}
\caption{Squared modulus of the wavefunction at $E = 1.8305$.}
\end{figure}
\begin{figure}[H]
\centering
\includegraphics[trim = 0 0 0 16 cm, clip = true]{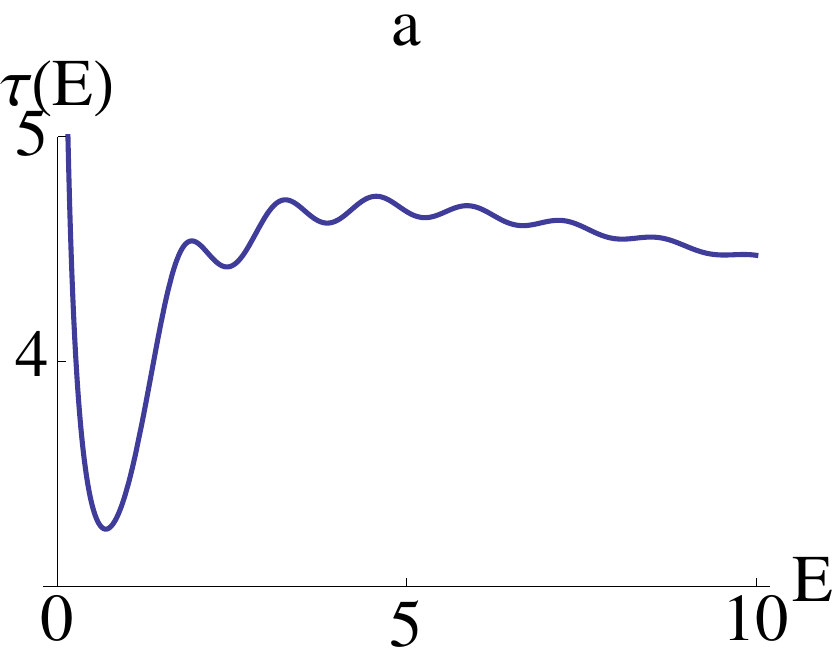}
\caption{Feeble oscillations in time-delay when $V(x < 0) = 0$ (Eq. (18)). Nevertheless, notice the closeness of $\epsilon_n$ and $E_n$.}
\end{figure}

\begin{figure}[H]
\centering
\includegraphics[trim = 0 0 0 16 cm, clip = true]{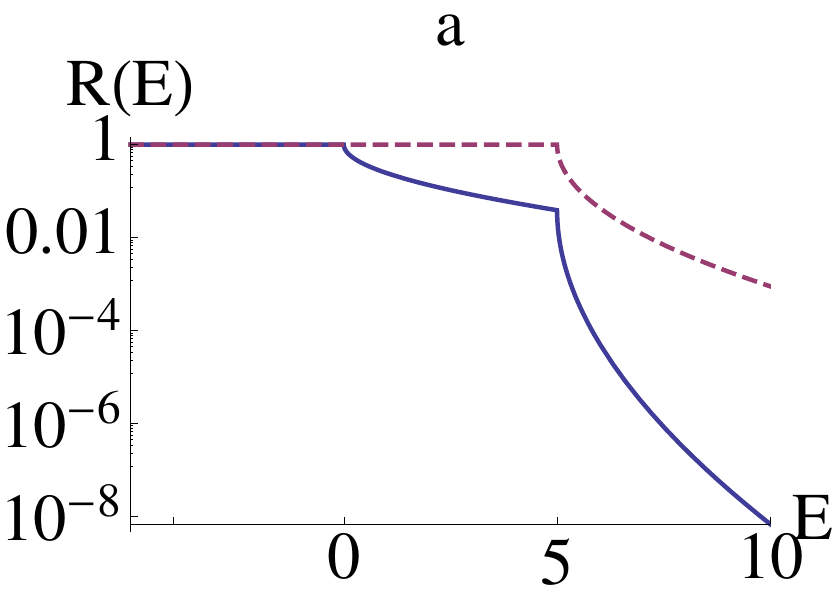}
\caption{Interesting three-piece reflectivity (R(E)) when $V(x \ge 0) = 0$, $V(x < 0) = V_1[1-e^{-2x/a}]$ (solid line). Dashed line is R(E) for one piece potential: $V_1[1-e^{-2x/a}]$.}
\end{figure}

\begin{table*} [t]
\caption{First five resonances in various systems. ${\cal E}_n = E_n - i\Gamma_n/2 ~(\Gamma_n > 0)$ are the poles of $r(E)$ and $\epsilon_n$ are the peak positions in time-delay, $\tau(E)$. We take $2m = 1 =\hbar^2$. Notice the general closeness of $E_n$ and $\epsilon_n $, excepting the case of {\bf P1}.}

\begin{ruledtabular}
\begin{tabular}{|c||c||c||c||c||c||c||c||c|}
		\hline
		Pn & Fig. & Eq. & Parameters & ${\cal E}_0 (\epsilon_0)$ & ${\cal E}_1 (\epsilon_1)$ & ${\cal E}_2 (\epsilon_2)$ & ${\cal E}_3 (\epsilon_3)$ & ${\cal E}_4 (\epsilon_4)$\\
		\hline
		\hline
		P1 & 2 & (16) & $a=1, b=1$ & $1.83 - 2.49 i$ & $5.39 - 6.09 i$ & $10.08 - 8.67 i$ & $14.33 - 12.50 i$ & $21.63 - 15.08 i$ \\
	& & &  $V_1=1, V_2 = 1$& (2.09) & (-) & (-) & (-) & (-)\\
	\hline
		P2 & 3 & (16) &  $a =1, b=5$ & $1.19 - 0.51 i$ & $2.87 - 0.84 i$ & $4.71 - 1.22 i$ & $6.83 - 1.56 i$ & $9.03 - 1.74 i $ \\
	 & & & $V_1=1, V_2 = 1$ & $(1.26)$ & $(2.98)$ & $(4.77)$ & $(6.96)$ & $(9.13)$\\
	\hline	
	P3 & 4 & (16) & $a =1, b=10$ & $0.77 - 0.22 i$ & $ 1.68 - 0.31 i$ & $2.58 - 0.39 i$ & $3.51 - 0.48 i$ & $4.46 - 0.57 i$ \\
	&  & & $V_1=1, V_2 = 1$ & $(0.77)$ & $(1.69)$ & $(2.58)$ & $(3.50)$ & $(4.45)$\\
	\hline		
	P4 & 5 & (16) & $a =1, b=5$ & $1.49 - 0.29 i$ & $ 3.29 - 0.44 i$ & $5.18 - 0.58 i$ & $7.18 - 0.72 i$ & $9.29 - 0.86 i$ \\
	&  & & $V_1=10, V_2 = 1$ & $(1.46)$ & $(3.28)$ & $(5.17)$ & $(7.19)$ & $(9.29)$\\
	\hline	
	P5 & 6 & (16) & $a =0.1, b=5$ & $1.53 - 0.11 i$ & $3.34 - 0.16 i$ & $5.23 - 0.21 i$ & $7.21 - 0.26 i$ & $ 9.30 - 0.31 i$ \\
	&  & & $V_1=1, V_2 = 1$ & $(1.51)$ & $(3.33)$ & $(5.22)$ & $(7.21)$ & $(9.29)$\\
	\hline		
	P6 & 7 & (16) &  $a =0.1, b=5$ & $0.46 - 0.02 i$ & $1.11 - 0.03 i$ & $ 1.85 - 0.05 i$ & $2.69 - 0.07 i$ & $3.60 - 0.08 i$ \\
	 & & & $V_1=1, V_2 = 0.1$ & $(0.45)$ & $(1.10)$ & $(1.84)$ & $(2.67)$ & $(3.59)$\\
	\hline	
	P7 & 10 & (18) &  $b=10$ & $1.76 - 0.91 i$ & $3.16 - 1.03 i$ & $4.51 - 1.13 i$ & $5.86 - 1.22 i$ & $7.22 - 1.29 i$ \\
	 & & & $V_1=0, V_2 = 2$ & $(1.89)$ & $(3.25)$ & $(4.55)$ & $(5.88)$ & $(7.25)$\\
	\hline	
\end{tabular}
\end{ruledtabular}
\end{table*}
The time-delay plots for the present rising exponential  potential obtained using the reflection amplitude (16) are shown in Figs 2-7. The position $\epsilon_n$ of  first five peaks/maxima in time delay and the corresponding complex energy poles, ${\cal E}_n$, of $r(E)$ (16) are compiled in Table 1. We have used $\hbar^2=1=2m$ in our calculations.

When the two-piece potential is anti-symmetric (see {\bf P1} in Table 1), similar to odd-parabolic potential [1], we find that the time-delay entails one maximum at $E=\epsilon_0\approx E_0$ and other resonances being broad (see the increasing value of $\Gamma_n$ in the Table 1) fail to cause a peak/maximum in time-delay. Also $\epsilon_0$ and $E_0$ differ quite a bit.
On the other hand, the smooth single piece rising potential $V(x)=-x^3$ [5] is rich in resonances.
However, in asymmetric cases of (2) ({\bf P2} to {\bf P6}), multiple resonances of varying quality occur wherein $\epsilon_n$ and $E_n$ are quite close.

{\bf P1} through {\bf P6} show increasing quality of resonances. More explicitly, the quality of resonances increases with increase in the parameter $b$ (compare cases {\bf P2} and {\bf P3}), decrease in parameter $a$ (see cases {\bf P2} and {\bf P5}), decrease in $V_2$ (cases {\bf P5} and {\bf P6}), and increase in $V_1$ (cases {\bf P2} and {\bf P4}).

Notice that the asymptotic scattering states (8) have a special feature of being energy independent. Therefore, regardless of whether energy is real or complex, they keep oscillating with reducing amplitude as $x\rightarrow -\infty$, without entailing spatial catastrophe even in a resonant eigenstate. Fig. 8 shows typical behavior of any resonant state of (2) and Fig. 9  illustrates the typical behavior of scattering state of (2) at any real positive energy. This also resolves the absence of catastrophe in resonant states of odd-parabolic potential where the asymptotic forms of the scattering states are
$\psi_{i,r}\sim e^{[\pm ix^2-(1 \mp i E)\log x]/2}$ [1]. Though these  are energy dependent, the dependence is suppressed as $x^2$ dominates over $\log x$ for large $x$. However, the two piece rising linear potential (discussed in Ref.[9]) displays spatial catastrophe  since its asymptotic form of the scattering states are energy dependent: $\psi_{i,r} \sim \phi(v(x))e^{\mp i \chi(v(x))}$, $\chi(v(x))\sim 2(-v(x))^{3/2}/3,$ as $x\rightarrow -\infty$, where $v(x)=\frac{2m}{\hbar^2} \frac{hx-E}{h^{2/3}}.$ In other words, when potential diverges,  the energy term in Schr{\"o}dinger equation becomes negligible. This causes no or very weak dependence on energy at asymptotically large distances. Thus,  we expect no catastrophe in the resonant states of $V(x)=x^{2n+1}$ or $V(x)=|x|x^{2n+1}$, $(n=1,2,3..)$, if scattering
from these potentials will be studied in future.

The case when $V_1=0$ is dealt separately wherein we get $V(x \le 0)=0$ and $V(x>0)=V_2[e^{2x/b}-1]$. The time-delay plot for this case though shows feeble oscillations, yet the agreement 
between $\epsilon_n$ and $E_n$ is excellent (see Table 1). Similar scenarios
have been presented [9] for the cases when the rising part of the potential is parabolic or linear.

The case when $V_2=0$ is the orthodox type of scattering potential which is devoid of resonances as there is only one real turning point that too at $E<0$. This potential has been obtained by cutting off $V(x)=V_1[1-e^{-2x/a}]$ on the right hand side such that $V(x \ge 0)=0$. The full potential has $R(E<V_1)=1$ [12], but in the cut-off case we get a  novel three-piece reflectivity in the domain $E \in (-\infty, 0],(0, V_1], (V_1, \infty)$. The time-delay (not shown here) is structureless
as there are no resonances.

Earlier, two-piece, one-dimensional, semi-infinite potentials have presented a surprising single deep minimum [13,14] in reflectivity.  Here, two-piece exponential potential shows surprising occurrence of resonances and unusual absence of catastrophe. In optics, [15] one investigates the wave propagation through various mediums and 
systems, the rising exponential potential presents a new possibility.

As stated in the beginning, many a times, the scattering from a rising potential can not be studied even numerically by integration of the Schr{\"o}dinger equation. In this regard, the analytic reflection amplitudes presented here are valuable.
The investigation of scattering from rising potentials have just begun, we hope that the rising exponential model presented
here will take this issue forward.

\end{document}